\documentclass[aps,prl,reprint, superscriptaddress]{revtex4-2}

\usepackage{graphicx}
\graphicspath{figures/}
\usepackage{dcolumn}
\usepackage{amssymb}
\usepackage{amsmath}
\usepackage{fancyhdr}
\usepackage{bm}
\usepackage{times}
\usepackage{mhchem}
\usepackage{ulem}
\usepackage[colorlinks,linkcolor=blue,anchorcolor=blue,citecolor=blue,urlcolor=blue]{hyperref}
\usepackage{braket}
\usepackage{dcolumn}
\usepackage{mathrsfs}
\usepackage{color}
\UseRawInputEncoding
\usepackage{appendix}
\usepackage{changes}
\usepackage{bibunits}

\UseRawInputEncoding
\begin{document}

\title{Mechanical Squeezed-Fock Qubit: Towards Quantum Weak-Force Sensing}
\author{Yi-Fan Qiao}
\affiliation{Ministry of Education Key Laboratory for Nonequilibrium Synthesis and Modulation of Condensed Matter, Shaanxi Province Key Laboratory of Quantum Information and Quantum Optoelectronic Devices, School of Physics, Xi'an Jiaotong University, Xi'an 710049, China}
\author{Jun-Hong An}
\affiliation{Key Laboratory of Quantum Theory and Applications of MoE, Lanzhou Center for Theoretical Physics, and Key Laboratory of Theoretical Physics of Gansu Province, Lanzhou University, Lanzhou 730000, China}

\author{Peng-Bo Li}
\email{lipengbo@mail.xjtu.edu.cn}
\affiliation{Ministry of Education Key Laboratory for Nonequilibrium Synthesis and Modulation of Condensed Matter, Shaanxi Province Key Laboratory of Quantum Information and Quantum Optoelectronic Devices, School of Physics, Xi'an Jiaotong University, Xi'an 710049, China}


\begin{abstract}
Mechanical qubits offer unique advantages over other qubit platforms, primarily in terms of coherence time and possibilities for enhanced sensing applications, but their potential is constrained by​ the inherently weak nonlinearities and small anharmonicity of nanomechanical resonators.
We propose to overcome this shortcoming by using squeezed Fock states of phonons in a parametrically driven nonlinear mechanical oscillator.
We find that, under two-phonon driving, squeezed Fock states become eigenstates of a Kerr-nonlinear mechanical oscillator, featuring an energy spectrum with exponentially enhanced and tunable anharmonicity, such that the transitions to higher energy states are exponentially suppressed. This enables us to encode the mechanical qubit within the ground and first excited squeezed Fock states of the driven mechanical oscillator. This kind of mechanical qubit is termed \textit{mechanical squeezed-Fock qubit}. We also show that our mechanical qubit can serve as a quantum sensor for weak forces, with its resulting sensitivity increased by at least one order of magnitude over that of traditional mechanical qubits. The proposed mechanical squeezed-Fock qubit provides a powerful quantum phonon platform for quantum sensing and information processing.
\end{abstract}
\maketitle
\begin{bibunit}[apsrev4-2]
\textit{Introduction.}---Mechanical resonators in the quantum regime represent a pivotal platform for quantum science and technologies~\cite{POOT2012273,Aspelmeyer2014,Moser2014,Norte2016,Hong2017,Satzinger2018,Riedinger2018,Dumur2021,Xu2022,Qiao2023,Zoepfl2023}, offering promising pathways toward scalable quantum computation and ultra-precise sensing~\cite{OckeloenKorppi2018,Tadokoro2021,Wollack2022,Chamberland2022,OckeloenKorppi2016,Fluehmann2019,Burd2019,Carney2021,Pachkawade2022,Bereyhi2022,Zhang2024}. Compared to their electromagnetic counterparts, mechanical resonators offer longer lifetimes, greater compactness \cite{Bachtold2022}, and the capability for direct coupling to different degrees of freedom~\cite{Bild2023,Szorkovszky2011,Kolkowitz2012,Faust2013,Puller2013,Rips2014,Jing2014,Li2016,Liao2016,Manenti2017,SanchezMunoz2018,Li2020,Wang2022,Beccari2022,Pan2024,Qiao2025}.
Such devices enable quantum squeezing of mechanical motion~\cite{Rabl2004,Wollman2015,LaHaye2004,Pirkkalainen2015,Marti2024,Zhou2024}, quantum transduction~\cite{Rabl2010,Mirhosseini2020,PhysRevLett.116.043601,Zhao2025}, and quantum entanglement~\cite{Jost2009,Palomaki2013,Lo2015,Lecocq2015,Chu2017,Kotler2021}.
In particular, mechanical resonators exhibiting nonlinearities have been viewed as intriguing candidates for the realization of mechanical qubits in quantum computation~\cite{Rips2013,Pistolesi2021,Yang2024}. Furthermore, their ability to couple to a variety of external weak forces, including gravitational
forces~\cite{Goryachev2014,Bonaldi2020,Bose2025}, magnetic forces, or others, enables quantum force sensing with substantially improved sensitivity~\cite{Caves1980,Bocko1996,Degen2017,Zhao2019,Asjad2023}.​

Traditionally, a mechanical qubit is encoded in the first two Fock states ($|0\rangle$ and $|1\rangle$) of a Kerr-nonlinear oscillator~\cite{Rips2013,Pistolesi2021,Yang2024}.
Coherent control via external fields requires the energy difference $\alpha$ between the $|0\rangle\rightarrow|1\rangle$ and $|1\rangle\rightarrow|2\rangle$ transitions (termed anharmonicity) to significantly exceed the phonon decoherence rate $\gamma_0$. This condition ensures negligible excitation of state $|2\rangle$ during quantum operations.
However, the development in mechanical qubits has been fundamentally limited by the inherent physical constraints of nanomechanical resonators---notably weak nonlinearities and insufficient anharmonicity---which hinder robust state engineering and coherent control. This limitation could be overcome by coupling mechanical resonators to other strongly
nonlinear quantum systems such as superconducting qubits~\cite{Kirchmair2013,Singh2020,Puebla2020,Kumar2021,Catalini2021,Yang2021,Samanta2023,Liu2024}. However, this may induce additional decoherence channels inherited from the auxiliary nonlinear quantum systems~\cite{Rips2013,Pistolesi2021,Yang2024}, which in turn limit the performance of mechanical qubits for quantum sensing and computation. Thus, a natural question arises: Can we find a new mechanism to boost mechanical oscillators with weak Kerr nonlinearities into high-fidelity qubits for quantum sensing with unprecedented sensitivity?

In this work, we provide an answer to the above question. We find that squeezed Fock states are the eigenstates of a Kerr-nonlinear mechanical oscillator under two-phonon driving, with a spectrum exhibiting exponentially enhanced and tunable anharmonicities. We show that the transitions to higher energy states are exponentially suppressed, and the dynamics of the system is confined to the  subspace spanned by the first and second squeezed Fock states. In this way, a mechanical qubit can be encoded in the first two squeezed-Fock phonon states of the driven Kerr-nonlinear oscillator, which is termed mechanical squeezed-Fock qubit.
Compared to traditional mechanical Fock qubits, this squeezed-Fock qubit proposal offers critical advantages: (i) Under weak Kerr nonlinearity (small $K$), it exponentially suppresses leakage to higher states by enhancing anharmonicity from $\alpha \sim K$  to $\alpha \sim Ke^{4r}$  (where $r$ is the two-phonon driving parameter), overcoming the inherent limitation of conventional designs. (ii) It maintains qubit validity across arbitrary regimes of Kerr nonlinearity---including $K > \gamma_0$, $K \sim \gamma_0$, and $K < \gamma_0$. This enables direct transformation of weakly anharmonic mechanical oscillators into robust qubits, dramatically simplifying experimental implementation by eliminating the need for strong intrinsic nonlinearity.

We employ this mechanical squeezed-Fock qubit for weak force sensing. In stark contrast to traditional mechanical Fock-qubit protocols~\cite{Degen2017,Pistolesi2021}, the sensitivity for weak force detection increases by one or two orders of magnitude.​
Our mechanical squeezed-Fock qubit provides a powerful new platform for quantum sensing and computation.

\begin{figure}
	\centering
	\includegraphics[width=0.45\textwidth]{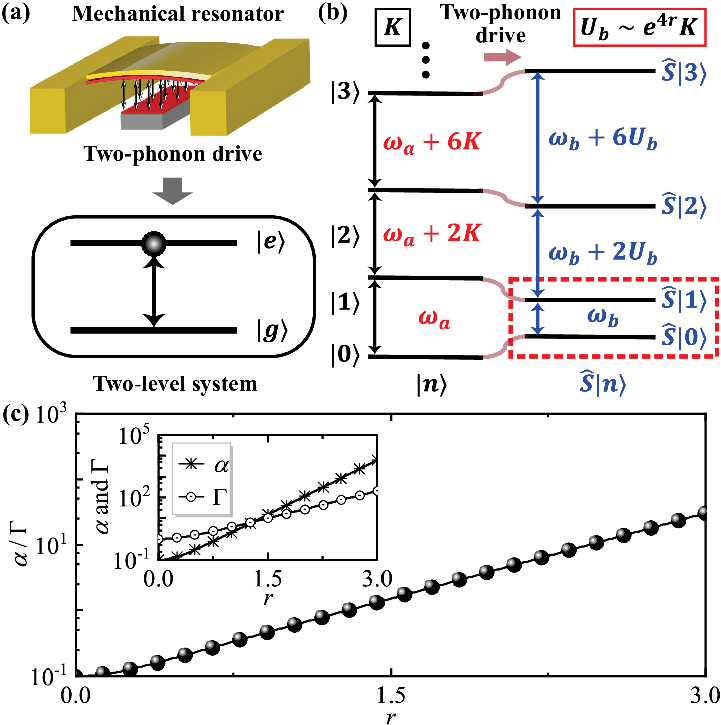}
	\caption{(a) Schematic of a doubly clamped mechanical resonator with weak Kerr nonlinearities. Applying a two-phonon drive, the system behaves as high-fidelity qubits rather than resonators.
		(b) Spectrum of a mechanical qubit without and with a two-phonon drive (not to scale). The nonlinear terms $K \hat{a}^\dag\hat{a}^\dag\hat{a}\hat{a}$ causes the harmonic spectrum of the mechanical resonator to become anharmonic, but the weak $K$ is not sufficient to select two distinct states as a qubit. The two-phonon drive makes the nonlinear intensity be enhanced exponentially with a scale of $\sim e^{4r}$, such that the two lowest states (red dashed square) can be chosen as a qubit.
		(c) The exponentially enhanced values of $\alpha$, $\Gamma$ and their ratio are shown on a logarithmic scale as functions of the parameter $r$.}
	\label{Fig1}
\end{figure}
\textit{Mechanical squeezed-Fock qubit.}---Conventionally, a mechanical resonator [as shown in Fig.~\ref{Fig1}(a)], whose eigenstates are the phonon Fock states $|n\rangle$ with eigenenergy $E_n$, is characterized by an equally spaced energy levels.
Introducing a Kerr nonlinearity can break this harmonicity, leading to unevenly spaced energy levels.
Since the Kerr nonlinear term commutes with the number operator, the eigenstates remain Fock states $|n\rangle$.
Consequently, the lowest two states can be selected to form an effective qubit. The Hamiltonian for this system is ($\hbar =1$)
\begin{equation}\label{Ori_H}
	\hat{H} = \omega_a \hat{a}^{\dagger}\hat{a} + K\hat{a}^{\dagger}\hat{a}^{\dagger}\hat{a}\hat{a},
\end{equation}
where $\omega_a$ is the resonator frequency, $\hat{a}^\dag$ ($\hat{a}$) is the phonon creation (annihilation) operator, and $K$ quantifies the strength of the phonon-phonon interaction.
The dynamics of the system is therefore governed by the Lindblad master equation:
\begin{equation}
	\frac{d \rho}{dt} = -i[\hat{H}, \rho] + \frac{\gamma_0}{2} \mathcal{D}_{\hat{a},\hat{a}^\dag} \rho,
\end{equation}
where $\rho$ is the system's density operator, and $\mathcal{D}_{\hat{A},\hat{B}}\rho \equiv 2\hat{A} \rho \hat{B} - \hat{B} \hat{A} \rho - \rho \hat{B} \hat{A}$.
Here, the anharmonicity $\alpha$ is introduced to assess whether the resonator can function as a mechanical qubit. This anharmonicity is defined as the energy difference between the $|1\rangle \rightarrow |2\rangle$ and $|0\rangle \rightarrow |1\rangle$ transitions: $\alpha_0 = 2K$.
For a mechanical qubit, the anharmonicity must significantly exceed the decoherence rate ($\alpha_0 \gg \gamma_0$). This ensures that the system's energy predominantly oscillates between the qubit states $|0\rangle$ and $|1\rangle$ with minimal leakage to $|2\rangle$ under the coherent quantum operations. However, achieving strong nonlinearity (large $K$) without auxiliary nonlinear quantum systems experimentally remains challenging.
For weak $K$, the small detuning between the $\{ |0\rangle, |1\rangle\}$ and $\{ |1\rangle, |2\rangle\}$  transitions leads to significant leakage to higher states, which compromises the validity of the two-level approximation. While leakage can be suppressed by using a sufficiently weak drive, this approach renders the qubit highly vulnerable to decoherence, thereby preventing its use in quantum tasks.
\begin{figure*}[t!]
	\centering
	\includegraphics[width=0.94\textwidth]{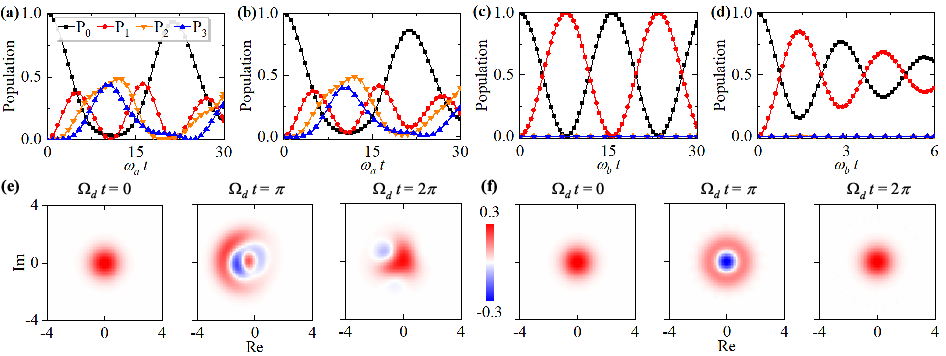}
	\caption{Validity of mechanical squeezed Fock qubit.
		(a) and (b) are the population $P_n$ of Fock states $|n\rangle$ in the absence and presence of decoherence. (c) and (d) are the population $P_n$ of squeezed Fock states $|n\rangle_S$ in the absence and presence of decoherence.
	    (e) and (f) are the Wigner functions of phonon modes at three points during a direct Fock qubit and squeezed Fock qubit Rabi oscillation, respectively. Here, the initial parameters $\gamma_0 = K$ in (b) and (d), $r = 1.5$ in (c), (d) and (f).}
	\label{Fig2}
\end{figure*}

We propose to achieve an exponentially enhanced anharmonicity by applying a two-phonon drive to the resonator. Specifically, we consider pumping the system with a periodic drive of frequency $2\omega_p$ and amplitude $\Omega_p$. In the rotating frame defined by the unitary transformation $\hat{U}_0(t) = \exp(-i\omega_p \hat{a}^{\dagger}\hat{a} t)$, the Hamiltonian (\ref{Ori_H}) becomes
\begin{equation}\label{H_tot}
	\hat{H} = \delta_a \hat{a}^{\dagger}\hat{a} + K\hat{a}^{\dagger}\hat{a}^{\dagger}\hat{a}\hat{a} + \frac{\Omega_p}{2}\left[e^{-i\theta}\hat{a}^2 + e^{i\theta}(\hat{a}^{\dagger})^2\right],
\end{equation}
where $\delta_a = \omega_a - \omega_p$ is the detuning and $\theta$ is the drive phase. The free and drive terms in Hamiltonian (\ref{H_tot}) can be diagonalized  via the Bogoliubov transformation $\hat{b} = \cosh r \cdot \hat{a} + e^{i\theta}\sinh r \cdot \hat{a}^{\dagger} = \hat{S}\hat{a} \hat{S}^\dagger $, with the squeezing operator $\hat{S} = \exp \left(r^* \hat{a}^2/2-r  \hat{a}^{\dagger 2}/2\right)$ and the squeezing parameter $r$ satisfies $\tanh (2r) = \Omega_p/\delta_a$.
Applying this transformation to the Kerr term, the Hamiltonian under the rotating-wave approximation is then given by~\cite{SM}
\begin{equation}
	\hat{H}_{\text{eff}} =\omega_b \hat{b}^\dag \hat{b} + U_b \hat{b}^{\dagger}\hat{b}^{\dagger}\hat{b}\hat{b},	
\end{equation}
where $\omega_b = \sqrt{\delta_a^2 - \Omega_p^2} + K(8\cosh^2 r \sinh^2 r + 4\sinh^4 r)$ represents the squeezed-resonator frequency and $U_b = [3\cosh(4r)+1]K/4 $ is the enhanced nonlinear strength. Obviously, the Bogoliubov transformation yields a new mechanical mode $\hat{b}$ with frequency $\omega_b$ and exponentially strengthened nonlinear interaction $U_b \sim e^{4r} K$.
In addition, the two-phonon drive transforms the system's eigenstates from Fock states $|n\rangle$ to squeezed Fock states $|n\rangle_S \equiv \hat{S}|n\rangle$~\cite{SM}, as illustrated in Fig.~\ref{Fig1}(b). Selecting the two lowest squeezed states $|0\rangle_S$ and $|1\rangle_S$ as qubit basis states yields the anharmonicity
\begin{equation}
	\alpha = 2U_b = \frac{3\cosh(4r) + 1}{2} K,
\end{equation}
which demonstrates an exponential enhancement proportional to $e^{4r}$.
The system dynamics is therefore governed by the master equation:
\begin{equation}
	\label{MQ}
	\begin{aligned}
		\frac{d \rho}{dt} =& -i[\hat{H}_{\text{eff}}, \rho] + \frac{\gamma_0}{2}(\mathcal{N}+1)\mathcal{D}_{\hat{b},\hat{b}^\dag} \rho+ \frac{\gamma_0}{2}\mathcal{N}\mathcal{D}_{\hat{b}^\dag,\hat{b}} \rho\\
		& -\frac{\gamma_0}{2}\mathcal{M}\mathcal{D}_{\hat{b},\hat{b}} \rho -\frac{\gamma_0}{2}\mathcal{M}^* \mathcal{D}_{\hat{b}^\dag,\hat{b}^\dag} \rho,
	\end{aligned}
\end{equation}
where $\mathcal{N} = \sinh^2 r$ and $\mathcal{M} = e^{-i\theta}\cosh r \sinh r$.
The effective decoherence rate is then enhanced to $\Gamma = \cosh (2r) \gamma_0$ \cite{SM}.
The first term governs coherent dynamics, while the second and third terms represent incoherent dissipation: $\mathcal{D}_{\hat{b},\hat{b}^\dag} \rho$ and $\mathcal{D}_{\hat{b}^\dag,\hat{b}} \rho$ correspond to energy emission and absorption processes, respectively. The final two terms describe two-phonon correlated processes induced by the two-phonon drive~\cite{Qin2018,Chen2021,Krstiifmmodeacutecelsecfi2024,Qin2024}. Here $\mathcal{N}$ and $\mathcal{M}$ are squeezing parameters with amplitude $r$ and phase angle $\theta$, where $\theta = 0$ and $\theta = \pi$ correspond to the squeezing of position and momentum space, respectively.

Note that the two-phonon drive not only exponentially enhances the anharmonicity but also amplifies noise.
However, the anharmonicity enhancement scales as $\sim e^{4r}$, significantly exceeding the noise amplification at a rate $\sim e^{2r}$. This contrast is illustrated in Fig.~\ref{Fig1}(c).
Initially assuming the decoherence rate $\gamma_0$ exceeds the bare anharmonicity $\alpha_0$ by an order of magnitude ($\gamma_0 \sim 10\alpha_0$), we observe both quantities grow exponentially with increasing $r$. Crucially, since the growth rate of $\alpha$ is $e^{2r}$ times that of $\Gamma$, $\alpha$ surpasses $\Gamma$ near $r=1.5$ and rapidly dominates at larger $r$. We further quantify this advantage through the phonon anharmonicity-decoherence ratio $\alpha/\Gamma$ versus $r$.
These results demonstrate that weak intrinsic nonlinearities ($\alpha_0 \ll \gamma_0$) can be exponentially enhanced via two-phonon driving. This enables robust encoding of a mechanical qubit in the two lowest squeezed Fock states $|0\rangle_S$ and $|1\rangle_S$.

To intuitively demonstrate the validity of the mechanical qubit encoded in the squeezed Fock states $|0\rangle_S$ and $|1\rangle_S$, Fig.~\ref{Fig2} compares the the populations as well as the Wigner function of phonons without (Fock states) and with (squeezed Fock states) the two-phonon drive, respectively.
Under coherent driving at rate $\Omega_d$ [Fig.~\ref{Fig2}(a) and (b) in the absence and presence of decoherence], the populations of Fock states $|2\rangle$ (blue triangles) and $|3\rangle$ (orange inverted triangles) exhibit significant temporal evolution.
In contrast, Fig.~\ref{Fig2}(c) demonstrates complete suppression of leakage to the squeezed Fock states $|2\rangle_S$ and $|3\rangle_S$, whose populations remain negligible. This confirms that the two-phonon drive confines the dynamics of the system to the qubit subspace, enabling predominant oscillations between $|0\rangle_S$ and $|1\rangle_S$.
When we take the decoherence into consideration, as shown in Fig.~\ref{Fig2}(d), we can obtain a damped oscillation between the two squeezed-Fock states $|0\rangle_S$ and $|1\rangle_S$.
From the perspective of the Wigner function, we can perform the qubit operations and visualize the resulting states, as shown in Fig.~\ref{Fig2}(e) and 2(f). We assume that the initial states are $|0\rangle$ and $|0\rangle_S$, corresponding to the cases without and with two-phonon drive, respectively.
For the case of the former, we can find that the mechanical system gradually evolves to the state with a sub-Poissonian distribution.
By contrast, the state under two-phonon drive evolves approximately
to $|1\rangle_S$ at $\Omega_d t = \pi$, 
and eventually returns to $|0\rangle_S$.
This further confirms that two-phonon driving enables weakly nonlinear mechanical resonators to behave as high-fidelity qubits rather than resonators.

\begin{figure}
	\centering
	\includegraphics[width=0.45\textwidth]{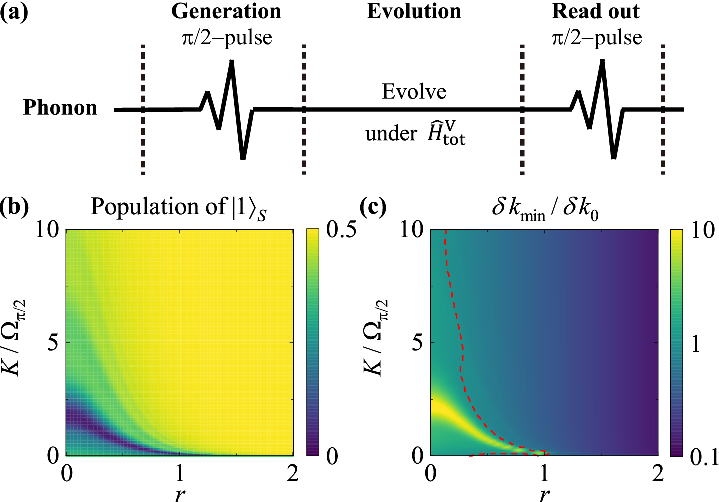}
	\caption{The validity of mechanical squeezed Fock qubit in the Ramsey sensing.
		(a) Flowchart of the Ramsey sensing.
		(b) The population of $|1\rangle_S$ obtained by the Ramsey sensing and (c) The relative sensitivity  $\delta k_{\text{min}}/\delta k_0$  versus nonlinear strength $K$ and parameter $r$. Here, $\Omega_{\pi/2}$ is the driving strength of the $\pi/2$ pulse and the initial parameter $\gamma_0 = \Omega_{\pi/2}$. }
	\label{Fig3}
\end{figure}
\textit{Sensing of the spring constant of restoring forces.}---We have thus established a mechanical qubit encoded in the squeezed Fock states $ |0\rangle_S$ and $| 1 \rangle_S$. We now consider utilizing this mechanical squeezed Fock qubit to sense the weak force field, with the coupling term given by $H_V=kx^2/2$, and $k$  the effective spring constant induced by the external signal.
Specifically, the qubit Hamiltonian is given by $\hat{H}_0=\omega_b \hat{\sigma}_z/2$, where ${\hat{\sigma}_x, \hat{\sigma}_y, \hat{\sigma}_z}$ denote the Pauli operators. The projection of signal Hamiltonian into the Hilbert space spanned by $\{ |0\rangle_S, \,|1\rangle_S\}$ can be expressed as $\hat{H}_V = \frac{1}{2} k x_0^2 e^{2r} \hat{\sigma}_z$~\cite{SM}. Here, $x_0 = 1/\sqrt{2m\omega_a}$ represents the zero-point fluctuation amplitude of the resonator with mass $m$. The total Hamiltonian is then
\begin{equation}
	\hat{H}_{\text{tot}}^{\text{V}} = \frac{1}{2} (\omega_b + \omega_V)\hat{\sigma}_z,
\end{equation}
where $\omega_V \equiv k x_0^2 e^{2r}$. The signal induces a frequency shift in the qubit, which can be directly measured via Ramsey interferometry [key steps are shown in Fig.~\ref{Fig3}(a)]~\cite{Degen2017}. Under dissipative decoherence, the encoding process is governed by the master equation:
\begin{equation}
	\begin{aligned}
		\frac{d \rho}{dt} = &-i[\hat{H}_{\text{tot}}^{\text{V}}, \rho] + \frac{\gamma_0}{2}(\mathcal{N}+1)\mathcal{D}_{\hat{\sigma}_-,\hat{\sigma}_+} \rho+ \frac{\gamma_0}{2}\mathcal{N}\mathcal{D}_{\hat{\sigma}_+,\hat{\sigma}_-} \rho\\
		& + \gamma_0 \mathcal{M} \hat{\sigma}_- \rho \hat{\sigma}_- - \gamma_0 \mathcal{M}^{*} \hat{\sigma}_+ \rho \hat{\sigma}_+,
	\end{aligned}
\end{equation}
where $\hat{\sigma}_- = \hat{\sigma}_+^\dag = |0\rangle_S \langle 1|$ are the ladder operators of qubit. This equation describes the dynamics of a two-level system under a squeezed thermal reservoir and we have used the condition: $\hat{\sigma}_- \hat{\sigma}_- = \hat{\sigma}_+\hat{\sigma}_+ = 0$. We assume the initial state is $|0\rangle_S$. After the initial $\pi/2$ pulse, free evolution and a second $\pi/2$ pulse, the system evolves into~\cite{Wang2017}
\begin{equation}
	\begin{aligned}
    \rho (t) &= \frac{1}{2}\{(1 + \text{Re}[c(t)]) \, |0\rangle_S \langle 0| + (1 - \text{Re}[c(t)]) \, |1\rangle_S \langle 1|]\\
    &+ [\kappa(1-|c(t)|^2) - i\, \text{Im}[c(t)]] |0\rangle_S\langle 1| + \text{H.c}\},
    \end{aligned}
\end{equation}
where $c(t) = \exp \{[-\gamma_0 \cosh(2r) /2 - i(\omega_b + \omega_V)]t \}$, $\kappa = 1/\cosh 2r$~\cite{SM}.
The probability $P_{|1\rangle_S}$ of measuring the qubit in the excited state is then given by
\begin{equation}
	    P_{|1\rangle_S} =\frac{1}{2} - \frac{1}{2} e^{-\gamma_0 \cosh(2r) t/2 }\cos[(\omega_b + \omega_V)t].
\end{equation}
This cosine function exhibits high sensitivity to small signal variations $\delta \omega_V$ at large evolution times $t$. In addition, the frequency $\omega_b + \omega_V$ can be determined from Ramsey fringes by measuring $P_{|1\rangle_S}$~\cite{Degen2017}. Assuming that the evolution time $t$ satisfies $(\omega_b + \omega_V)t = m\pi/2$ with a large integer $m=1,\,3,\,5 \cdots$, we have $P_{|1\rangle_S}=1/2$ , which corresponds to the points of maximum slope~\cite{Degen2017}.

We simulate the $P_{|1\rangle_S}$ during the Ramsey measurement process under different parameters in Fig.~\ref{Fig3}(b). In the absence of two-phonon drive, i.e., $r = 0$, weak Kerr nonlinearity $K$ causes significant leakage to higher states, leading to $P_{|1\rangle_S}<0.5$. Here, the observed oscillations arise from the interference between primary and leakage evolution paths.
With the introduction of the two-phonon drive, we can find that both energy leakage and parasitic oscillations are exponentially suppressed.
In addition, the quantity of interest is the difference $\delta P_{|1\rangle_S}$ between the probability measured before and after a perturbation of the qubit, which can be expressed as~\cite{SM}
\begin{equation}
	\delta P_{|1\rangle_S} = \pm \frac{1}{2}e^{- \gamma_0 \cosh (2r)t/2}x_0^2e^{2r} \delta k t  ,
\end{equation}
where $\delta k$ denotes the spring constant variation. We can find the difference $\delta P_{|1\rangle_S} $ grows linearly with evolution time $t$. It indicates that longer $t$ means more sensitivity to changes in the external field (higher resolution), but it is also subject to the qubit decoherence.

We next estimate the fundamental sensitivity through signal-to-noise ratio (SNR) analysis, which is defined as~\cite{Degen2017}
\begin{equation}
    \text{SNR} =  \frac{\delta P_{|1\rangle_S}}{\sigma_p} = e^{- \gamma_0 \cosh (2r)t/2}x_0^2e^{2r} \delta k C\sqrt{N} t,
\end{equation}
where $\sigma_p = 1/2C\sqrt{N}$ is the total noise with the number of experiment $N = T/(t+t_m)$, $T$ is the total available measurement time and $t_m$ is the extra time needed to initialize, manipulate, and read out the sensor, $C$ represents the efficiency parameter of readout.
We take the value $C=1$ and $t_m=0$.
The sensitivity is then defined as the minimum detectable signal $\delta k_{\text{min}}$ that yields unit SNR for an integration time of $T = 1$ s~\cite{Pistolesi2021}. Its minimal over the encoding time is
\begin{equation}
	\delta k_{\text{min}} = \frac{\sqrt{\gamma_0 \cosh(2r) e } }{x_0^2 e^{2r}} \approx \frac{\sqrt{\gamma_0 e} }{x_0^2 e^{r}},
\end{equation}
which is obtained in the condition of $d \delta k_{\text{min}} / dt = 0 $, i.e., $t=1/[\gamma_0 \cosh(2r)]$. We have shown the relative sensitivity $\delta k_{\text{min}}/\delta k_0$ in Fig.~\ref{Fig3}(c), where $\delta k_0= \sqrt{\gamma_0 e}/x_0^2$ is the standard sensitivity of $kx^2/2$ sensing by the mechanical Fock qubit~\cite{SM}. Here, the red dashed line indicates the reference $\delta k_{\text{min}}/\delta k_0 = 1$.  We can clearly see that $\delta k_{\text{min}}/\delta k_0$ decreases as $r$ increases. The sensitivity is exponentially promoted by a factor $\sim e^r$ compared to conventional mechanical Fock qubits, which indicates that our proposal provides higher precision for quantum sensing.

\textit{Experimental feasibility.}---To explore the experimental feasibility of our proposal, we here discuss the relevant parameters.
We consider a mechanical resonator with frequency $\omega_a/2\pi=600$ MHz and decoherence $\gamma_0/2\pi=3$~kHz. For a finite temperature of 10 mK, the average thermal phonon numbers $\overline{n} \approx 0.00595 \ll 1$ and can be ignored.
Assuming a weak Kerr nonlinear  strength $K/2\pi=3$~kHz, we apply a two-phonon drive to the mechanical mode satisfying $\Omega_p /\delta_a = \tanh (2r) $ with $r=1.5$. In this case, the anharmonicity and decoherence rate are then enhanced to $\sim 1.8$~MHz and $\sim 13.5$~kHz, where we have $\Gamma \ll \alpha$, such that the mechanical resonator can be seen as a two-level system with a frequency $\omega_b/2\pi =4.2$~MHz. Mechanical resonators with weak nonlinearity have been reported in nanomechanical systems such as  cantilevers, membranes or carbon nanotubes~\cite{Kim2013,Djorwe2014,ArellanoCastro2014,Ella2015,Lai2022,Venkatachalam2023}. In addition, our proposed scheme is also applicable to the systems whose nonlinearities are introduced by coupling to two energy level systems such as superconducting qubits~\cite{Liu2010,Ramos2013,Wollman2015,Chu2017}.

We next consider using this mechanical squeezed Fock qubit to sense weak force field $kx^2/2$. For a typical example of carbon nanotube resonators with frequency $\omega_a$, the mass is $m=10^{-21}$ kg, we have the zero-point fluctuation $x_0 = 4 \times 10^{-12}$ m.
Under a decoherence rate of $\gamma_0/2\pi=3$ kHz, the calculated sensitivity reaches $\delta k_{\text{min}} = 4.71 \times 10^{-10}$ $\text{N}\cdot \text{m}^{-1}/\text{Hz}^{1/2}$. This represents an order-of-magnitude improvement over the undriven case with identical parameters ($\delta k_0 = 2.11 \times 10^{-9}$ $\text{N}\cdot \text{m}^{-1}/\text{Hz}^{1/2}$), and the enhancement becomes more pronounced at larger values of $r$.

\textit{Conclusion.}---In conclusion, we have shown that, by applying a two-phonon drive to mechanical resonators with weak Kerr nonlinearities, one can obtain an exponentially enhanced and tunable anharmonicity.
Though the decoherence is also expeonentialy enhanced, the main benefit is that the enhancement of the former is much greater than that of the latter.
This scheme enables us to encode a mechanical qubit in the first two squeezed-Fock phonon states.
The mechanical qubit can couple to a variety of forces and thus be used to sense weak signals. Compared to the traditional mechanical Fock qubit, the sensitivity using the squeezed-Fock qubit has been exponentially promoted.
The proposed squeezed-Fock state encoding is general and can be applied to other bosonic systems like photons and magnons.

\begin{acknowledgments}
P.B.L. is supported by the National Natural Science Foundation of China under Grants No. W2411002 and No. 12375018. J.H.A. is supported by the National Natural Science Foundation of China (Grants No. 12275109 and No. 12247101) and the Innovation Program for Quantum Science and Technology of China (Grant No. 2023ZD0300904).
\end{acknowledgments}

%
\end{bibunit}

\clearpage
\pagestyle{empty}
\onecolumngrid
\vspace*{10pt}
\renewcommand{\theequation}{S\arabic{equation}}
\setcounter{equation}{0}
\begin{bibunit}[apsrev4-2]
\begin{center}
	\large \textbf{Supplemental Material for ``Mechanical squeezed-Fock qubit: Towards Quantum Weak-Force Sensing"}
\end{center}

\section{abstract}
	This Supplementary Material (SM) provides additional details to support and deepen the understanding presented in the main text.
	Section I presents the specific derivation of the enhanced nonlinearity and the master equation resulting from the application of a two-phonon drive to the mechanical resonator.
	Section II details the protocol for sensing the spring constant associated with the restoring force and derives the sensitivity using the mechanical squeezed-Fock qubit via the Ramsey measurement method. For comparison, the sensitivity achieved with a conventional mechanical Fock qubit under the same Ramsey measurement protocol is also derived. Additionally, as a supplementary application, we discuss the use of the mechanical squeezed-Fock qubit for sensing a static force of the form $Fx$
	using the Rabi measurement method.



\section{Exponentially enhanced nonlinearity by two-phonon drive}
\subsection{Effective Hamiltonian}
In this section, we provide a detailed derivation demonstrating the exponential enhancement of the nonlinearity in a mechanical resonator subjected to a two-phonon drive. For a mechanical resonator, the Hamiltonian is given by
\begin{equation}\label{1}
	\hat{H} = \omega_a \hat{a}^\dagger \hat{a} ,
\end{equation}
where $\omega_a$ is the frequency of the resonator, and the operators $\hat{a}^{\dagger}$ and $\hat{a}$ are the phonon creation and annihilation operators.
It is obvious that its eigenstates and eigenvalues are the Fock states $|n \rangle$ and $ n\omega_a$ as $\hat{a}^\dag \hat{a} |n \rangle = n |n \rangle $. In the presence of a Kerr nonlinear, the Hamiltonian (\ref{1}) becomes to
\begin{equation}\label{2}
	\hat{H} = \omega_a \hat{a}^\dagger \hat{a} + K \hat{a}^{\dagger}\hat{a}^{\dagger}\hat{a}\hat{a},
\end{equation}
where $K$ is the strength of the phonon-phonon interaction.
One can calculate that $ K \hat{a}^{\dagger}\hat{a}^{\dagger}\hat{a}\hat{a} |n \rangle = Kn(n-1)|n \rangle $. It indicates that the eigenstates of the Hamiltonian (\ref{2}) are still Fock states $|n\rangle$, but the Kerr nonlinear term could break the harmonicity and modulate the eigenfrequencies of the mechanical resonator.
We now consider the case of a mechanical resonator possessing weak Kerr nonlinearity and pumped by a two-phonon drive, the Hamiltonian is then described by
\begin{equation}\label{ori}
	\hat{H}=\omega_a \hat{a}^{\dagger}\hat{a} + K \hat{a}^{\dagger} \hat{a}^{\dagger} \hat{a} \hat{a} + \Omega_p \cos (2\omega_pt) (e^{-i\theta}\hat{a} + e^{i\theta}\hat {a}^{\dagger})^2,
\end{equation}
where $2\omega_p$ is the frequency of the drive. The amplitude and phase of the drive are denoted by $\Omega_p$ and $\theta$.
Transforming into the rotating frame via the unitary operator $\hat{U}_0(t) = e^{-i\omega_p \hat{a}^{\dagger}\hat{a} t}$, the Hamiltonian becomes:	
\begin{equation}
	\hat{H}=\delta_a \hat{a}^{\dagger}\hat{a} + K\hat{a}^{\dagger}\hat{a}^{\dagger}\hat{a}\hat{a} + \frac{\Omega_p}{2}[e^{-i\theta}\hat{a}^2 + e^{i\theta}(\hat {a}^{\dagger})^2],
\end{equation}
where $\delta_a = \omega_a - \omega_p$.
The free and drive terms $\hat{H}_0 = \delta_a \hat{a}^{\dagger}\hat{a} + \Omega_p[e^{-i\theta}\hat{a}^2 + e^{i\theta}(\hat {a}^{\dagger})^2]/2$
can be diagonalized by the Bogoliubov squeezing transformation $\hat{b} = \hat{S} \hat{a} \hat{S}^\dag$ with $\hat{S} = \exp \left(r^* \hat{a}^2/2-r  \hat{a}^{\dagger 2}/2 \right)$:
\begin{equation}
	\begin{aligned}
		\hat{b} &= \cosh r\, \hat{a} + e^{i\theta} \sinh r\, \hat{a}^{\dagger}, \\
		\hat{b}^{\dagger} &= e^{-i\theta} \sinh r\, \hat{a} + \cosh r\, \hat{a}^{\dagger}.
	\end{aligned}
\end{equation}
The inverse transformation expresses the original modes as:
\begin{equation}
	\begin{aligned}
		\hat{a} &= \cosh r\, \hat{b} - e^{i\theta} \sinh r\, \hat{b}^{\dagger}, \\
		\hat{a}^{\dagger} &= \cosh r\, \hat{b}^{\dagger} - e^{-i\theta} \sinh r\, \hat{b}.
	\end{aligned}
\end{equation}
Substituting the Bogoliubov transformation into $\hat{H}_0$, we obtain the expanded form:
\begin{equation}
	\begin{aligned}
		\hat{H}_0 &= [\frac{\Omega_p}{2} (\cosh^2 r + \sinh^2 r) - \delta_a \cosh r \sinh r ](e^{i\theta}\hat{b}^{\dagger}\hat{b}^{\dagger} + e^{-i\theta}\hat{b}\hat{b}) \\
		&+ [\delta_a (\cosh^2 r + \sinh^2 r) - 2\Omega_p \cosh r \sinh r] \hat{b}^{\dagger}\hat{b} + \delta_a \sinh^2 r - \Omega_p \cosh r \sinh r .
	\end{aligned}
\end{equation}
Setting the coefficient of the squeezing terms to zero: $[\Omega_p (\cosh^2 r + \sinh^2 r)]/2 - \delta_a \cosh r \sinh r = 0$, we obtain the squeezing parameters: $\tanh (2r) = \Omega_p / \delta_a$, $\sinh {2r} = \Omega_p / \sqrt{\delta_a^2 - \Omega_p^2}$ and $\cosh {2r} = \delta_a / \sqrt{\delta_a^2 - \Omega_p^2}$, the diagonalized Hamiltonian $\hat{H}_0 = \delta_b \hat{b}^\dag \hat{b}$ with $\delta_b = \delta_a (\cosh^2 r + \sinh^2 r) - 2\Omega_p \cosh r \sinh r = \sqrt{\delta_a^2 - \Omega_p^2}$. Constant terms have been omitted in the final expression.
Expanding the Kerr term $K\hat{a}^{\dagger}\hat{a}^{\dagger}\hat{a}\hat{a}$ in the squeezed basis, the total Hamiltonian then reduces to:
\begin{equation}
	\begin{aligned}
		\hat{H} &=\omega_b \hat{b}^\dag \hat{b} + \frac{3\cosh(4r)+1}{4} K \hat{b}^{\dagger}\hat{b}^{\dagger} \hat{b}\hat{b} \\
		&+K[\frac{1}{4}\sinh^2 (2r)( e^{i2\theta} \hat{b}^{\dagger}\hat{b}^{\dagger}\hat{b}^{\dagger}\hat{b}^{\dagger}+ e^{-i2\theta}\hat{b}\hat{b}\hat{b}\hat{b}) - \frac{1}{2}\sinh (4r) (e^{i\theta}\hat{b}^{\dagger}\hat{b}^{\dagger} \hat{b}^{\dagger}\hat{b} +e^{-i\theta}\hat{b}^{\dagger}\hat{b}\hat{b}\hat{b}) \\
		&+ \frac{1}{2} \sinh (2r)(3\cosh (2r)-2)(e^{i\theta}\hat{b}^{\dagger}\hat{b}^{\dagger}+e^{-i\theta}\hat{b}\hat{b})  ],
	\end{aligned}
\end{equation}
where the renormalized frequency is given by $\omega_b = \delta_b + K (8\cosh^2 r \sinh^2 r + 4\sinh^4 r)$.
Performing the unitary transformation $\hat{U}_1(t) = e^{-i\omega_b \hat{b}^{\dagger}\hat{b} t}$ to move into the rotating frame yields the Hamiltonian:
\begin{equation}
	\begin{aligned}
		\hat{H}&= K \frac{3\cosh(4r)+1}{4} \hat{b}^{\dagger}\hat{b}^{\dagger}\hat{b}\hat{b} + \frac{\sinh (2r)(3\cosh (2r)-2)}{2} e^{i2\omega_b t}(e^{i\theta}\hat{b}^{\dagger}\hat{b}^{\dagger}+e^{-i\theta}\hat{b}\hat{b}] \\
		&+K[\frac{\sinh^2 (2r)}{4}e^{i4\omega_b t}( e^{i2\theta}\hat{b}^{\dagger}\hat{b}^{\dagger}\hat{b}^{\dagger}\hat{b}^{\dagger}+ e^{-i2\theta}\hat{b}\hat{b}\hat{b}\hat{b}) - \frac{\sinh (4r)}{2}e^{i2\omega_b t} (e^{i\theta}\hat{b}^{\dagger}\hat{b}^{\dagger} \hat{b}^{\dagger}\hat{b} +e^{-i\theta}\hat{b}^{\dagger}\hat{b}\hat{b}\hat{b}).
	\end{aligned}
\end{equation}
Under the conditions
\begin{equation}
	\begin{aligned}
		\frac{K \sinh^2 (2r)}{4} &\ll 4\omega_b, \
		\frac{K \sinh (4r)}{2} &\ll 2\omega_b, \
		\frac{K \sinh (2r)(3\cosh (2r) - 2)}{2} &\ll 2\omega_b,
	\end{aligned}
\end{equation}
the rapidly oscillating terms can be neglected (rotating wave approximation). The Hamiltonian in the interaction picture then simplifies to
\begin{equation}
	\hat{H}_I = U_b \hat{b}^{\dagger}\hat{b}^{\dagger}\hat{b}\hat{b},
\end{equation}
where $U_b = K[3\cosh(4r) + 1]/4$. Thus, we obtain a transformed mechanical mode $\hat{b}$ with an enhanced nonlinear interaction characterized by $U_b \hat{b}^{\dagger}\hat{b}^{\dagger}\hat{b}\hat{b}$. The effective Hamiltonian in the Schr\"odinger picture is therefore
\begin{equation}\label{eff}
	\hat{H}_{\text{eff}} = \omega_b \hat{b}^\dagger \hat{b} + U_b \hat{b}^{\dagger}\hat{b}^{\dagger}\hat{b}\hat{b}.
\end{equation}
We can find that the effective Hamiltonian (\ref{eff}) has the same form of Hamiltonian (\ref{2}), so the eigenstates of effective Hamiltonian in the squeezing representation are also the Fock states. We label the eigenstats as $|n\rangle_S$, where the symbol $S$ denotes the squeezing representation. The effective Hamiltonian can be expressed as the original mechanical mode $\hat{a}$ by the transformation $\hat{b} = \hat{S} \hat{a} \hat{S}^\dag$:
\begin{equation}
	\hat{H}_{\text{eff}} = \hat{S} \hat{H}_{\text{ori}} \hat{S}^\dag.
\end{equation}
Here, $\hat{H}_{\text{ori}} = \omega_b \hat{a}^\dagger \hat{a} + U_b \hat{a}^{\dagger}\hat{a}^{\dagger}\hat{a}\hat{a}$, whose eigenstates in the Schrodinger's representation are also Fock states $|n\rangle$ and eigenvalues is denoted $E_n^b$. So we have
\begin{equation}
	\hat{S} \hat{H}_{\text{ori}}|n\rangle = \hat{S} \hat{H}_{\text{ori}} \hat{S}^\dag \hat{S}|n\rangle = \hat{H}_{\text{eff}} \hat{S}|n\rangle = E_n^b \hat{S}|n\rangle.
\end{equation}
We can find in the Schrodinger's representation the egienstates of effective Hamiltonian (\ref{eff}) are the squeezed Fock states: $|n\rangle_S \equiv \hat{S}|n\rangle$.

\subsection{Master equation}
Without two-phonon drive, the system dynamic is governed by the Lindbland  master equation
\begin{equation}
	\frac{d \rho}{dt} = -i[H, \rho] + \frac{\gamma_0}{2} \mathcal{L}[a]
\end{equation}
where $\mathcal{L}[a]\rho =2\hat{a} \rho \hat{a}^\dag - \hat{a}^\dag \hat{a} \rho -\rho \hat{a}^\dag \hat{a}$ and $\gamma_0$ is the decoherence rate. The form of $\hat{H}$ here is Hamiltonian (\ref{2}). After applying the two-phonon drive, we perform the same transformation on the master equation as on the Hamiltonian and obtain
\begin{equation}
	\begin{aligned}
		\frac{d \rho}{dt} = -i[H, \rho] &+ \frac{\gamma_0}{2} (\cosh^2 r \hat{b} \rho \hat{b}^{\dagger} -  e^{-i\theta}\cosh r \sinh r \hat{b} \rho \hat{b} - e^{i\theta}\sinh r \cosh r \hat{b}^{\dagger} \rho \hat{b}^{\dagger} + \sinh^2 r \hat{b}^{\dagger} \rho \hat{b}) \\
		&- (\cosh^2 r \hat{b}^{\dagger}\hat{b}\rho - e^{i\theta}\cosh r \sinh r \hat{b}^{\dagger}\hat{b}^{\dagger}\rho - e^{-i\theta}\sinh r \cosh r \hat{b}\hat{b} \rho + \sinh^2 r \hat{b} \hat{b}^{\dagger} \rho) \\
		&- (\cosh^2 r \rho\hat{b}^{\dagger}\hat{b} - e^{i\theta}\cosh r \sinh r \rho\hat{b}^{\dagger}\hat{b}^{\dagger} - e^{-i\theta}\sinh r \cosh r \rho\hat{b}\hat{b}  + \sinh^2 r \rho\hat{b} \hat{b}^{\dagger}  ) .
	\end{aligned}
\end{equation}
The last three terms can be expanded and recombined as
\begin{equation}\label{ME}
	\begin{aligned}
		\frac{d \rho}{dt} =& -i[H, \rho] + \frac{\gamma_0}{2}(\mathcal{N}+1)\mathcal{L}[\hat{b}]\rho + \frac{\gamma_0}{2}\mathcal{N}\mathcal{L}[\hat{b}^\dag]\rho \\
		& -\frac{\gamma_0}{2}\mathcal{M} (2\hat{b} \rho \hat{b} - \hat{b}\hat{b} \rho - \rho\hat{b}\hat{b}) -\frac{\gamma_0}{2}\mathcal{M}^{*} (2\hat{b}^{\dagger} \rho \hat{b}^{\dagger} - \hat{b}^{\dagger}\hat{b}^{\dagger} \rho - \rho\hat{b}^{\dagger}\hat{b}^{\dagger})
	\end{aligned}
\end{equation}
where $\mathcal{N}=\sinh^2 r$ and $\mathcal{M} = e^{-i\theta}\cosh r\sinh r$.
To prove the validity of the transformation and approximation, we respectively simulated in FIG.~\ref{Fig1} the temporal evolution of the system population and $Q$-function using the original Hamiltonian (\ref{ori}) and the effective Hamiltonian (\ref{eff}). Comparison of  FIG.~\ref{Fig1}(a) and \ref{Fig1}(b), as well as \ref{Fig1}(c) and \ref{Fig1}(d), reveals nearly identical population distributions and $Q$-functions between the original and effective Hamiltonian, demonstrating the validity of the approximation.
\begin{figure*}[t!]
	\centering
	\includegraphics[width=0.94\textwidth]{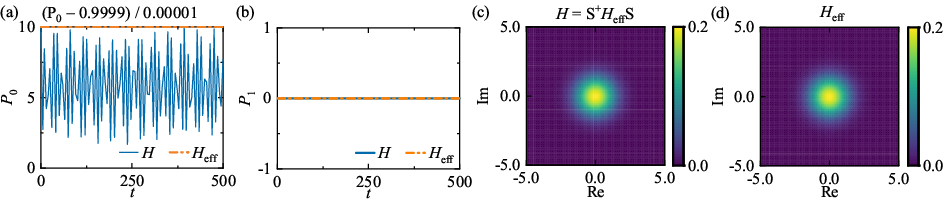}
	\caption{Validity of the transformation and approximation.
		(a) Population $P_0$ of state $|0\rangle$ under the original Hamiltonian (\ref{ori}) and state $|0\rangle_S$ under the effective Hamiltonian (\ref{eff}) as a function of time $t$.
		(b) Population $P_1$ of state $|1\rangle$ with original Hamiltonian (\ref{ori}) and state $|1\rangle_S \equiv \hat{S}|1\rangle$ with effective Hamiltonian (\ref{eff}) as a function of time $t$.
		(c) $Q$-function calculated using the original Hamiltonian (\ref{ori}).
		(d) $Q$-function  calculated using the effective Hamiltonian (\ref{eff}). (a) vs. (b) and (c) vs. (d) show excellent agreement in both population dynamics and $Q$-functions for the original and effective Hamiltonians, validating the approximation. The parameter is $r=1$.}
	\label{Fig1}
\end{figure*}

\section{Sensing with the mechanical squeezed qubits}
Consider a mechanical qubit formed by the two states $|0\rangle_S$ and $|1\rangle_S$. Obviously, any operator $\hat{O}$ can be projected in the Hilbert space spanned by $\{|g\rangle,\, |e\rangle \}$ via a linear combination of the unit matrix $\hat{\sigma}_0$ and the three Pauli operators $\{ \hat{\sigma}_x, \hat{\sigma}_y, \hat{\sigma}_z \}$:
\begin{equation}
	\hat{O} = a_0 \hat{\sigma}_0 + a_x \hat{\sigma}_x + a_y\hat{\sigma}_y + a_z\hat{\sigma}_z,
\end{equation}
whose matrice forms are defined by
\begin{equation}
	\begin{aligned}
		&|g\rangle = \begin{pmatrix} 0 \\ 1 \end{pmatrix}, \quad
		|e\rangle = \begin{pmatrix} 1 \\ 0 \end{pmatrix}, \quad
		\hat{I} = \begin{pmatrix} 1 & 0 \\ 0 & 1 \end{pmatrix}, \quad  \\
		\hat{\sigma}_x = \begin{pmatrix} 0 & 1 \\ 1 & 0 \end{pmatrix}, \quad
		\hat{\sigma}_y &= \begin{pmatrix} 0 & -i \\ i & 0 \end{pmatrix}, \quad
		\hat{\sigma}_z = \begin{pmatrix} 1 & 0 \\ 0 & -1 \end{pmatrix}, \quad
		\hat{\sigma}_+ = \begin{pmatrix} 0 & 1 \\ 0 & 0 \end{pmatrix}, \quad
		\hat{\sigma}_- = \begin{pmatrix} 0 & 0 \\ 1 & 0 \end{pmatrix} .
	\end{aligned}	
\end{equation}
In addition, each matrix element of the operator $\hat{O}$ projected in the Hilbert space spanned by $\{|0\rangle_S,\,|1\rangle_S \}$ can be calculated by $\hat{O}_{mn} = \langle m | \hat{O} | n \rangle$. The free Hamiltonian of the qubit is described by
\begin{equation}
	\hat{H}_0 = \omega_b |1\rangle_S\langle 1|=\frac{1}{2} \omega_b|1\rangle_S\langle 1| - \frac{1}{2} \omega_b|0\rangle_S\langle 0|= \frac{1}{2} \omega_b \hat{\sigma}_z,
\end{equation}
where $\omega_b$ is the transition energy between the states.
As the mechanical oscillator can be strongly coupled to forces, we next consider using this two-level system as a sensor qubit to sense the weak force field.

\subsection{Sensing of the spring constant of restoring forces with the squezeed-Fock qubit}
Consider the coupling term is described by
\begin{equation}
	\hat{V} =\frac{1}{2} k \hat{x}^2 ,
\end{equation}
where $k$ is the effective spring constant induced by the external signal and $\hat{x} =x_0 (\hat{a} + \hat{a}^\dag)$ is the displacement operator with $x_0 = 1/\sqrt{2m\omega_a} $, $m$ denotes the resonator mass. We first project this Hamiltonian into the Hilbert space spanned by $ \{ |0 \rangle_S,\, |1 \rangle_S \}$:
\begin{equation}
	\hat{H}_V = \frac{1}{2} k \hat{x}^2 \equiv \frac{1}{2} k x_0^2 (\hat{a} + \hat{a}^\dag)^2 \equiv \frac{1}{2} k x_0^2 e^{2r} (\hat{b} + \hat{b}^\dag)^2 = \frac{1}{2} k x_0^2 e^{2r} \begin{pmatrix} 3 & 0 \\ 0 & 1 \end{pmatrix} = \frac{1}{2} k x_0^2 e^{2r} \hat{\sigma}_z.
\end{equation}
Here, we have use the projection relation $\hat{b} (\hat{b}^\dag) \rightarrow \hat{\sigma}_- (\hat{\sigma}_+)$. The total Hamiltonian is then given by
\begin{equation}
	\hat{H}_{\text{tot}}^{\text{V}} = \hat{H}_0 + \hat{H}_V = \frac{1}{2} \omega_b \hat{\sigma}_z +\frac{1}{2} k x_0^2 e^{2r} \hat{\sigma}_z = \frac{1}{2} (\omega_b + \omega_V) \hat{\sigma}_z,
\end{equation}
where $\omega_V \equiv k x_0^2 e^{2r}$. The signal induces a frequency shift in the qubit, which can be directly measured via Ramsey interferometry measurement~\cite{Degen2017}. The Ramsey interferometry measurement mainly contains the following five steps.
\begin{enumerate}
	\item \textbf{Initialize}: The sensor is initialized into the qubit basis state $|0\rangle_S$;
	\item \textbf{Transform}: The sensor is next transformed into a desired initial sensing state. Ramsey interferometry measurement uses a $\pi/2$ pulse to transform the sensor into a superposition state. Here we assume that the pulse causes the quantum state to rotate counterclockwise by 90 degrees around the $y$-axis, whose Hamiltonian can be described by $\hat{H}_{\text{drive}} = -\Omega_{\pi/2} \hat{\sigma}_y/2$ and the matrix is
	\begin{equation}
		R_y(\frac{\pi}{2})= \frac{1}{\sqrt{2}} \begin{pmatrix} 1 & -1 \\ 1 & 1 \end{pmatrix}.
	\end{equation}
	The density operator is then given by
	\begin{equation}
		\rho_0 = |\psi_0\rangle_S \langle \psi_0| =R_y(\frac{\pi}{2}) |0\rangle_S \langle 0 | R_y^\dag(\frac{\pi}{2}) =\frac{1}{2}(|0\rangle_S \langle 0| + |1\rangle_S \langle 1| + |0\rangle_S \langle 1| +|1\rangle_S \langle 0|).
	\end{equation}
	The corresponding superposition state is $|\psi_0\rangle_S = (|0\rangle_S + |1\rangle_S)/\sqrt{2}$.
	\item \textbf{Evolve}: The sensor evolves under the Hamiltonian $\hat{H}_{\text{tot}}^{\text{V}}$ for a time $t$. Under dissipative decoherence, the dynamics of the encoding process is governed by the master equation
	\begin{equation} \label{me}
		\frac{d \rho}{dt} = -i[\hat{H}_{\text{tot}}^{\text{V}}, \rho] + \frac{\gamma_0 \cosh^2 r}{2}\mathcal{L}[\hat{\sigma}_-]\rho + \frac{\gamma_0 \sinh^2 r}{2}\mathcal{L}[\hat{\sigma}_+]\rho + \gamma_0 \cosh r \sinh r \hat{\sigma}_- \rho \hat{\sigma}_- + \gamma_0 \cosh r \sinh r \hat{\sigma}_+ \rho \hat{\sigma}_+.
	\end{equation}
	This master equation (\ref{me}) is derived by projecting the equation (\ref{ME}) into the qubit subspace, with the parameter $\theta$ set to $\pi$. It describes the evolution of a two-level system coupled to a squeezed thermal reservoir. Here, we have utilized the property $\hat{\sigma}_- \hat{\sigma}_- = \hat{\sigma}_+\hat{\sigma}_+ = 0$. The final two terms in the equation represent correlated two-quantum jump processes arising from the two-phonon drive. These terms encode the phase-sensitive quantum noise correlations characteristic of squeezing, directly coupling coherences during the dynamical evolution. Since they do not influence population dynamics, we neglect them in the subsequent population calculations for convenience. Expanding the master equation, we obtain:
	\begin{equation}
		\begin{aligned}
			\frac{d \rho}{dt} =& -i \frac{\omega_b +\omega_V}{2} [(|1\rangle_S \langle 1 | -|0\rangle_S \langle 0 |)\rho - \rho (|1\rangle_S \langle 1 | -|0\rangle_S \langle 0 |) ] \\
			& + \frac{\gamma \cosh^2 r}{2}[ 2 |0\rangle_S \langle 1 | \rho |1\rangle_S \langle 0 | - |1\rangle_S \langle 0 |0\rangle_S \langle 1 | \rho - \rho |1\rangle_S \langle 0 |0\rangle_S \langle 1 |] \\
			& + \frac{\gamma \sinh^2 r}{2}[ 2 |1\rangle_S \langle 0 | \rho |0\rangle_S \langle 1 | - |0\rangle_S \langle 1 |1\rangle_S \langle 0 | \rho - \rho |0\rangle_S \langle 1 |1\rangle_S \langle 0 |].
		\end{aligned}
	\end{equation}
	The evolution of each density matrix element is calculated using $d\rho_{mn}/dt =\langle m | d\rho /dt |n \rangle_S$, yielding the following equations:
	\begin{equation}
		\begin{aligned}
			\frac{d \rho_{00}}{dt} &=-\gamma_0 \sinh^2 r \rho_{00} + \gamma_0 \cosh^2 r \rho_{11}, \\
			\frac{d \rho_{11}}{dt} &= \gamma_0 \sinh^2 r \rho_{00} - \gamma_0 \cosh^2 r \rho_{11}, \\
			\frac{d \rho_{01}}{dt} &= i(\omega_b + \omega_V) \rho_{01} - \frac{\gamma_0 \cosh (2r)}{2} \rho_{01}, \\
			\frac{d \rho_{10}}{dt} &=-i(\omega_b + \omega_V) \rho_{10} - \frac{\gamma_0 \cosh (2r)}{2} \rho_{10}.
		\end{aligned}
	\end{equation}
	These differential equations admit the solutions:
	\begin{align}
		\rho_{gg}(t) &= \frac{1}{2} \left( 1 + \kappa \right) - \frac{\kappa}{2} e^{-\Gamma t}, \\
		\rho_{ee}(t) &= \frac{1}{2} \left( 1 - \kappa \right) + \frac{\kappa}{2} e^{-\Gamma t}, \\
		\rho_{ge}(t) &= \frac{1}{2} \exp\left( -i\omega t - \frac{\Gamma}{2} t \right), \\
		\rho_{eg}(t) &= \frac{1}{2} \exp\left( i\omega t - \frac{\Gamma}{2} t \right),
	\end{align}
	where $\Gamma = \gamma_0 \cosh 2r$, $\kappa = 1/\cosh 2r$ and $\omega = \omega_b + \omega_V$.
	\item \textbf{Transform}: After evolution time $t$, using a second $\pi/2$ pulse  converts the state back to the measurable state. The transformed density matrix $\rho = R_y(\frac{\pi}{2}) \rho(t) R_y^\dagger(\frac{\pi}{2})$ has elements:
	\begin{align}
		\rho'_{gg} &= \frac{1}{2} \left[( 1 - e^{-\Gamma t/2} \cos(\omega t) \right], \\
		\rho'_{ee} &= \frac{1}{2} \left[( 1 + e^{-\Gamma t/2} \cos(\omega t) \right], \\
		\rho'_{ge} &= \frac{1}{2} \left[ \kappa (1 - e^{-\Gamma t}) - i e^{-\Gamma t/2} \sin(\omega t) \right], \\
		\rho'_{eg} &= \frac{1}{2} \left[ \kappa (1 - e^{-\Gamma t}) + i e^{-\Gamma t/2} \sin(\omega t) \right].
	\end{align}
	The complete form of the final density operator is then give by
	\begin{equation}\label{37}
		\rho (t) = \frac{1}{2}\{ (1 + \text{Re}[c(t)]) \, |0\rangle_S \langle 0| + (1 - \text{Re}[c(t)]) \, |1\rangle_S \langle 1|+ \{[\kappa(1-|c(t)|^2) - \text{i}\, \text{Im}[c(t)]] |0\rangle_S\langle 1| + \text{H.c}]\} \},
	\end{equation}
	where $c(t) = \exp \{[-\gamma_0 \cosh(2r) /2 - i(\omega_b + \omega_V)]t \}$.
	\item \textbf{Readout}: The transition probability is then given by
	\begin{equation}\label{38}
		P_{|1\rangle_S} =\frac{1}{2} - \frac{1}{2} e^{-\gamma_0 \cosh(2r) t/2 }\cos[(\omega_b + \omega_V)t].
	\end{equation}
\end{enumerate}
Note that the results presented here have incorporated decoherence effects. The observed Ramsey fringes in $P_{|1\rangle_S}$, oscillating at frequency $\omega_b + \omega_V$, directly yield a measurement of the energy splitting.

The fundamental objective of quantum sensing is the detection of exceptionally weak physical signals, including minute magnetic fields, electric fields, mechanical forces, and accelerations. Such signals typically lie beyond the detection threshold of classical methodologies. To achieve high-precision measurement, an optimal strategy involves quantifying the deviation $\delta P= P - P_0$ of the transition probability $P$ from a deliberately selected reference value $P_0$, rather than measuring its absolute magnitude~\cite{Degen2017}. This reference value $P_0$ is designated as the bias point of the measurement, establishing the operational baseline for the detection process.
The bias point $P_0$ can be configured through two primary approaches. (1) Correspondence to known signal values: Achieved when the external signal (e.g., magnetic field) equals a calibrated reference value; (2) Parameter adjustment: Implemented by experimentally tuning controllable parameters in the system Hamiltonian to actively stabilize the measurement at $P_0$.
As demonstrated by Equation (\ref{38}), the bias point achieving maximal sensitivity in Ramsey interferometry measurement corresponds to the maximum slope of the cosine function, where the transition probability $P=0.5$.
This condition is satisfied when $(\omega_b + \omega_V)t = k\pi/2$ for odd integers $k = 1, 3, 5, \cdots$.
Considering a small perturbation $\delta \omega_V$ in the target signal, the variation in the transition probability relative to the reference point $P_0$ is given by
\begin{equation}
	\begin{aligned}
		\delta P_{|1\rangle_S} = P_{|1\rangle_S} - P_0 &= \frac{1}{2} - \frac{1}{2} e^{-\gamma_0 \cosh(2r) t/2 }\cos[(\omega_b + \omega_V + \delta \omega_V)t] - \frac{1}{2} \\
		&= \pm \frac{1}{2} e^{-\gamma_0 \cosh(2r) t/2 } \sin (\delta \omega_Vt) \\
		&\approx \pm \frac{1}{2} e^{-\gamma_0 \cosh(2r) t/2 } \cdot \delta \omega_Vt \\
		&= \pm \frac{1}{2}e^{- \gamma_0 \cosh (2r)t/2}x_0^2e^{2r} \delta k t  ,
	\end{aligned}
\end{equation}
where the Taylor expansion $\sin(\delta \omega_V t) \approx \delta \omega_V t$ is applied for small arguments, and $\delta k$ denotes the variation in the spring constant. To quantitatively characterize the measurement sensitivity, we first define the signal-to-noise ratio (SNR). The sensitivity $\delta k_{\text{min}}$ is correspondingly defined as the minimum detectable signal variation $\delta k$ per unit time. For quantum sensing experiments, the SNR is expressed as~\cite{Degen2017}
\begin{equation}
	\text{SNR} = \frac{\delta P_{\ket{1}_S}}{\sigma_P}
	= \frac{1}{2} e^{-\gamma_0 \cosh(2r) t/2} x_0^2 e^{2r} \delta k t \cdot 2C\sqrt{N},
	\label{eq:snr_definition}
\end{equation}
where $\sigma_P = 1/(2C\sqrt{N})$ denotes the total measurement noise, $N$ represents the number of experimental repetitions, and $C$ is the readout efficiency parameter. The ideal case $C=1$ corresponds to vanishing classical readout noise~\cite{Degen2017}. For simplicity, we adopt $C=1$ in subsequent analyses.
Additionally, the number of measurements $N$ relates to the total available time $T$ as $N = T/(t + t_m)$, where $t_m$ represents the overhead time for sensor initialization, manipulation, and readout. Assuming this overhead is negligible ($t_m \approx 0$), the SNR expression simplifies to
\begin{equation}
	\text{SNR} = e^{-\gamma_0 \cosh(2r) t/2} x_0^2 e^{2r} \delta k \sqrt{tT}.
	\label{eq:simplified_snr}
\end{equation}
The sensitivity $\delta k_{\text{min}}$, defined as the minimum detectable signal per unit time at unit SNR ($\text{SNR} = 1$) for $T = 1$ s integration time, is given by
\begin{equation}
	\delta k_{\text{min}} = \frac{e^{\gamma_0 \cosh(2r) t/2}}{x_0^2 e^{2r} \sqrt{t}}.
	\label{eq:deltak_min}
\end{equation}
Minimizing this expression with respect to the encoding time $t$ yields
\begin{equation}\label{sen}
	\delta k_{\text{min}}^{\text{opt}} = \frac{\sqrt{\gamma_0 \cosh(2r) e } }{x_0^2 e^{2r}} \approx \frac{\sqrt{\gamma_0 e} }{x_0^2 e^{r}},
\end{equation}
where the optimum occurs at $t = 1/[\gamma_0 \cosh(2r)]$, satisfying the minimization condition $d(\delta k_{\text{min}})/dt = 0$.

In addition to obtaining the sensitivity via the slope detection method used in the main text, we can also achieve the same result using the methods of error propagation and quantum Fisher information. We start from the density matrix Eq. (\ref{37}) after the second $\pi/2$ pulse. The measurement of the excited state population $\hat{O}=\hat{\sigma}_+\hat{\sigma}_-$ yields two values and corresponding probabilities:
\begin{equation}\label{44}
	\begin{aligned}
		1,\, P_1 &= \frac{1-\text{Re}[c(t)]}{2}, \\
		0,\, P_0 &= \frac{1+\text{Re}[c(t)]}{2}.
	\end{aligned}
\end{equation}
Its expectation value and variance are calculated as
\begin{equation}
	\begin{aligned}
		\bar{\hat{O}} &= 1 \times P_1 + 0 \times P_0 = \frac{1-\text{Re}[c(t)]}{2}, \\
		\Delta \hat{O} & = \sqrt{\bar{\hat{O^2}} - \bar{\hat{O}}^2} =\frac{\sqrt{1-\text{Re}[c(t)]^2}}{2},
	\end{aligned}
\end{equation}
where we used $\hat{O}^2 = \hat{O}$. Repeating the experiment for duration $T$ yields $N=T/t$ measurement results. According to Cram\'{e}r-Rao theorem, the measurement error for $\hat{O}$ is
\begin{equation}
	\delta \hat{O} = \frac{\sqrt{1-\text{Re}[c(t)]^2}}{2\sqrt{T/t}}.
\end{equation}
Utilizing the error propagation formula then gives
\begin{equation}
	\delta k = \frac{\delta \hat{O}}{|d\bar{\hat{O}}/dk|} = \frac{\sqrt{1-e^{\gamma_0 \cosh (2r)t}\cos^2[(\omega_b+\omega_V)t]}}{\sqrt{T/t}\sin[(\omega_b+\omega_V)t]tx_0^2e^{2r}}
\end{equation}
Choosing $\omega_b$ properly such that $\sin [(\omega_b+\omega_V)t]t = 1$ yields
\begin{equation}\label{48}
	\delta k_\text{min} = \frac{e^{\gamma_0\cosh (2r)t/2}}{\sqrt{Tt}x_0^2e^{2r}}.
\end{equation}
Its minimum over the encoding time is
\begin{equation}\label{sen1}
	\delta k_{\text{min}}^{\text{opt}} = \frac{\sqrt{\gamma_0 \cosh(2r) e } }{\sqrt{T}x_0^2 e^{2r}},
\end{equation}
obtained when $d(\delta k_{\text{min}})/dt = 0$, i.e., at $t = 1/[\gamma_0 \cosh(2r)]$. This result is identical to Eq. (\ref{sen}) calculated using the slope detection method.

\subsection{Sensing of the spring constant of restoring forces with Fock qubit}
For comparison, we present here the sensitivity to sense restoring force using the conventional mechanical Fock qubit.
Without two-phonon drive, the total Hamiltonian of the system is given by
\begin{equation}
	\hat{H} = \omega_a \hat{a}^\dagger \hat{a} + K \hat{a}^{\dagger}\hat{a}^{\dagger}\hat{a}\hat{a},
\end{equation}
whose eigensates are Fock states $|n\rangle$. We assume the nonlinear strength $K$ is so large that we can regard the lowest two states $|0\rangle$ and $|1\rangle$ as a qubit sensor. Next we turn to the Hilbert space spanned by $\{|0\rangle,\,|1\rangle\}$, and the Hamiltonian of the Fock qubit is then given by
\begin{equation}
	\hat{H}_0^{*} = \omega_a |1\rangle \langle 1|=\frac{1}{2} \omega_a|1\rangle\langle 1| - \frac{1}{2} \omega_a|0\rangle\langle 0|= \frac{1}{2} \omega_a \hat{\sigma}_z^*,
\end{equation}
where $\omega_a$ is the transition energy between the qubit states and $\hat{\sigma}_z^* = |1\rangle\langle 1| - |0\rangle\langle 0|$. Projecting the signal Hamiltonian $\hat{H}_V = k\hat{x}^2/2$ to the subspace of $\{ |0\rangle, \,|1\rangle \}$, we have
\begin{equation}
	\hat{H}_V = \frac{1}{2} k \hat{x}^2 \equiv \frac{1}{2} k x_0 (\hat{a} + \hat{a}^\dag)^2 = \frac{1}{2} k x_0 \begin{pmatrix} 3 & 0 \\ 0 & 1 \end{pmatrix} = \frac{1}{2} k x_0 \hat{\sigma}_z^*.
\end{equation}
Here, we have use the projection relation $\hat{a} (\hat{a}^\dag) \rightarrow \hat{\sigma}_-^* (\hat{\sigma}_+)^* \equiv |0\rangle \langle 1 | (|1\rangle \langle 0 |)$. The Hamiltonian then becomes
\begin{equation}
	\hat{H}_{\text{tot}}^{\text{V}*} = \hat{H}_0^* + \hat{H}_V = \frac{1}{2} \omega_a \hat{\sigma}_z^* +\frac{1}{2} k x_0 \hat{\sigma}_z^* = \frac{1}{2} (\omega_a + \omega_V) \hat{\sigma}_z^*,
\end{equation}
where $\omega_V \equiv k x_0$. In this case, the encoding process in the presence of the dissipative decoherence is governed by
\begin{equation}
	\frac{d\hat{\rho}}{dt} = -i[\hat{H}_{\text{tot}}^{\text{V}*}, \hat{\rho}] + \frac{\gamma_0 }{2}\mathcal{L}[\hat{\sigma}_-^*]\rho.
\end{equation}
We repeat the Ramsay measurement process from the previous section and assume that the initial state is $|0\rangle$.
After the first $\pi/2$ pulse, the density operator is then given by
\begin{equation}
	\rho_0^* = |\psi_0^*\rangle \langle \psi_0^*|=\frac{1}{2}(|0\rangle \langle 0| + |1\rangle \langle 1| + |0\rangle \langle 1| +|1\rangle \langle 0|),
\end{equation}
where $|\psi_0^*\rangle = (|0\rangle  + |1\rangle)/\sqrt{2}$. After evolving under the Hamiltonian $\hat{H}_{\text{tot}}^{\text{V}*}$ for a time $t$, the evolved state is
\begin{equation}
	\rho^*(t) = \frac{1}{2}\left[ (2 - |c(t)|^2)\ket{0}\bra{0} + |c(t)|^2\ket{1}\bra{1} + \left( c(t)^*\ket{0}\bra{1} + c(t)\ket{1}\bra{0} \right) \right],
\end{equation}
where $c(t) = e^{-\gamma_0 t/2 - i(\omega_a + \omega_V)t}$. Next, the second \(\pi/2\) pulse converts \(\rho(t)\) into~\cite{Wang2017}
\begin{equation}
	\rho (t) = \frac{1}{2}\{(1 + \text{Re}[c(t)]) \, |0\rangle_S \langle 0| + (1 - \text{Re}[c(t)]) \, |1\rangle_S \langle 1|+ [(1-|c(t)|^2 - \text{i}\, \text{Im}[c(t)] )|0\rangle_S\langle 1| + \text{H.c}]\}.
\end{equation}
The transition probability is then given by
\begin{equation}
	P_{|1\rangle_S}^* =\frac{1}{2} - \frac{1}{2} e^{-\gamma_0 t/2 }\cos[(\omega_a + \omega_V)t].
\end{equation}
Repeating the same steps as in the previous section to obtain the minimum detectable signal variance from the SNR, we can obtain the sensitivity $\delta k_0$ sensing via a mechanical Fock qubit $\delta k_0$
\begin{equation}
	\delta k_0= \frac{\sqrt{\gamma_0 e}}{x_0^2},
\end{equation}
where the optimum occurs at $t = 1/\gamma_0$. Compared to Eq.~(\ref{sen}), we can find the sensitivity sensing by a mechanical squeezed qubit is exponentially promoted by a factor $\sim e^r$.

\subsection{Sensing of the static forces}
As a supplementary application, mechanical squeezed qubits can also be employed for static force sensing. We assume the coupling between the mechanical sensor qubit and the force is described by $\hat{H}_F = F\hat{x}$, where $F$ is the external force and $\hat{x} = x_0 (\hat{a} + \hat{a}^\dag)$ is the displacement operator induced by the force $F$ acting on the mechanical oscillator, with the amplitude of the zero-point quantum fluctuation given by $x_0 =1/\sqrt{2m\omega_a}$. We first project this Hamiltonian into the Hilbert space spanned by $ \{ |0 \rangle_S,\, |1 \rangle_S \}$, resulting in
\begin{equation}
	\hat{H}_F = F \hat{x} = F x_0 (\hat{a} + \hat{a}^\dag) = F x_0 e^{r} (\hat{b} + \hat{b}^\dag) = F x_0 e^{r} \begin{pmatrix} 0 & 1 \\ 1 & 0 \end{pmatrix} = F x_0 e^{r} \hat{\sigma}_x,
\end{equation}
where we have chosen the phase $\theta = \pi$ for convenience, corresponding to momentum space squeezing. This yields the total Hamiltonian
\begin{equation}
	\hat{H}_{\text{tot}}^{\text{F}} = \hat{H}_0 + \hat{H}_F = \frac{1}{2}\hbar \omega_b \hat{\sigma}_z + \frac{1}{2}\hbar \omega_F \hat{\sigma}x,
\end{equation}
where $\omega_F \equiv 2x_0e^{r}F/\hbar$. Unlike the case of sensing restoring forces, which causes a frequency shift of the qubit, the transverse signal here explicitly introduces transitions between qubit states~\cite{Degen2017}. We next solve for the transition probability. Assuming the qubit evolves under $\hat{H}_{\text{tot}}^{\text{F}}$ for time $t$, the final state becomes
\begin{equation}
	|\psi \rangle = C_1(t) |1\rangle_S + C_0(t) |0\rangle_S,
\end{equation}
where $C_1(t)$ and $C_0(t)$ are complex coefficients. According to the Schr\"odinger equation $i\hbar \frac{d}{dt}|\psi \rangle = \hat{H}|\psi \rangle$, the evolution of the coefficients is governed by
\begin{equation}
	\begin{aligned}
		2\text{i} \frac{d}{dt}{C}_0(t) &= \omega_b C_0(t) + \omega_F C_1(t) , \\
		2\text{i} \frac{d}{dt}{C}_1(t) &= \omega_F C_0(t) + \omega_b C_1(t) .
	\end{aligned}
\end{equation}
The formal solution to this system of equations is
\begin{equation}
	\begin{aligned}
		{C}_0(t) &= \left[\cos \left(\frac{\omega_R t}{2}\right) + \text{i} \frac{\omega_b}{\omega_R} \sin \left(\frac{\omega_R t}{2}\right)\right] C_0(0) + \text{i} \frac{\omega_F}{\omega_R} \sin \left(\frac{\omega_R t}{2}\right) C_1(0) , \\
		{C}_1(t) &= \text{i} \frac{\omega_F}{\omega_R} \sin \left(\frac{\omega_R t}{2}\right) C_0(0) + \left[\cos \left(\frac{\omega_R t}{2}\right) - \text{i} \frac{\omega_b}{\omega_R} \sin \left(\frac{\omega_R t}{2}\right)\right] C_1(0) ,
	\end{aligned}
\end{equation}
where the effective Rabi frequency is $\omega_R = \sqrt{\omega_b^2 + \omega_F^2}$, and $C_0(0)$, $C_1(0)$ denote the initial population distribution at $t=0$. Assuming the initial state $|\psi_0 \rangle = |0\rangle_S$, the transition probability to state $|1\rangle_S$ at time $t$ is
\begin{equation}
	P{|1\rangle_S} = |C_1 (t)|^2 = \frac{\omega_F^2}{\omega_R^2} \sin^2 \left(\frac{\omega_R}{2}t\right) = \frac{\omega_F^2}{2\omega_R^2} \left[1-\cos (\omega_Rt)\right].
\end{equation}
This shows that a slight change in force $F$ induces transitions between states $|1\rangle_S$ and $|0\rangle_S$, manifesting as an increased transition rate.

Next, we calculate the effect of a small force variation $\delta F$ on the transition probability $P_{|1\rangle_S}$. We choose the bias point at maximum slope, where $P_0 = \omega_F^2/2\omega_R^2$ with $\omega_Rt=\pi/2+k\pi , (k=1,2,3...)$. Assuming the perturbation $\delta F$ has negligible influence on $\omega_F^2/\omega_R^2$, we focus primarily on its effect on the sinusoidal term. For sufficiently long measurement time $t$, the difference in transition probability $\delta P_{|1\rangle_S} = P_{|1\rangle_S}-P_0$ under $\delta F$ is, to first order,
\begin{equation}
	\delta P_{|1\rangle_S} = P_{|1\rangle_S}-P_0 = -\frac{\omega_F^2}{2\omega_R^2}\cos (\omega_R^{'} t) \approx -\frac{\omega_F^2}{2\omega_R^2}\cos \left(\sqrt{\omega_b^2 + \omega_F^2 + 2\omega_F \delta \omega_F }t\right) \approx -\frac{\omega_F^2}{2\omega_R^2}\cos \left(\omega_R t + \frac{\omega_F \delta \omega_F}{\omega_R}t\right)
\end{equation}
where $\omega_R^{'} = \sqrt{\omega_b^2 + \omega_F^{'2}}$ with $\omega_F^{'} = \omega_F + \delta \omega_F$, and we used the Taylor expansion $\sqrt{1 + x} \approx 1+ x/2$. Given $\omega_Rt=\pi/2+k\pi$, this simplifies to
\begin{equation}
	\delta P_{|1\rangle_S} =  \frac{\omega_F^2}{2\omega_R^2} \sin \left(\frac{\omega_F \delta \omega_F}{\omega_R}t\right) \approx  \frac{\omega_F^2}{2\omega_R^2} \times \frac{\omega_F \delta \omega_F}{\omega_R}t =  \frac{\omega_F^3}{\omega_R^3} \frac{ x_0 e^{r} \delta F}{\hbar}t.
\end{equation}
The probability shift $\delta P_{|1\rangle_S}$ is exponentially enhanced by the squeezing factor $e^r$ in our scheme. It is important to note that these results are for the ideal case without decoherence. In practice, decoherence and relaxation introduce errors that reduce the observable probability with increasing sensing time $t$~\cite{Degen2017}:
\begin{equation}
	\delta P_{\text{obs}}(t) = \delta P_{|1\rangle_S} (t) e^{-\chi (t)}
\end{equation}
where $\chi (t)$ is a phenomenological decoherence function capturing the relevant noise processes. As established earlier, the effective decoherence rate is exponentially enhanced to $\gamma_0 \cosh (2r)$ in the presence of two-phonon drive. Consequently, the decoherence function takes the specific form $\chi (t) = \gamma_0 \cosh (2r) t/2$. The signal-to-noise ratio (SNR) is then
\begin{equation}
	\text{SNR} = \frac{\delta P_{\text{obs}}}{\sigma_p} =  \frac{\omega_F^3}{\omega_R^3} \frac{ x_0 e^{r} \delta F}{\hbar}t e^{- \gamma_0 \cosh (2r) t/2} 2 C \frac{\sqrt{T}}{\sqrt{t+t_m}},
\end{equation}
leading to the minimum detectable signal
\begin{equation}
	\delta F_\text{min} = \frac{\omega_R^3}{\omega_F^3} \frac{\hbar e^{\gamma_0 \cosh (2r) t/2 }}{2x_0 e^{r}\sqrt{t}},
\end{equation}
where we also set $C=1$ and $t_m \approx 0$ as before. Minimizing this expression with respect to the encoding time $t$ yields
\begin{equation}\label{senF0}
	\delta F_{\text{min}}^{\text{opt}} = \frac{\omega_R^3}{\omega_F^3} \frac{\hbar\sqrt{e\gamma_0 \cosh(2r)}}{2x_0e^r} \approx \frac{\omega_R^3}{\omega_F^3}\frac{\hbar\sqrt{e\gamma_0}}{2x_0},
\end{equation}
where the optimum occurs at $t = 1/[\gamma_0 \cosh(2r)]$, satisfying $d(\delta F_{\text{min}})/dt = 0$. To achieve a large signal, it is preferable to have $\omega_F$ comparable to or larger than $\omega_b$ (thus $\omega_R$). The sensitivity (\ref{senF0}) ultimately matches the value reported in Ref.~~\cite{Pistolesi2021}. However, it should be emphasized that our scheme requires only a two-phonon drive, whereas the scheme in Ref.~~\cite{Pistolesi2021} necessitates coupling to auxiliary systems to induce sufficient anharmonicity. This additional coupling may introduce extra decoherence channels from auxiliary nonlinear quantum systems, potentially limiting the performance of mechanical qubits in quantum sensing and computation applications.
Asumming $\omega_R^3/\omega_F^3 \sim 1$  , we plot in FIG. (\ref{Fig2}) the population at the bias point and sensitivity changing with the nonlinear strength $K$ and parameter $r$. We can find that for the weak nonlinear strength $K$, the leakage of population can be exponentially suppressed.
\begin{figure*}[t!]
	\centering
	\includegraphics[width=0.47\textwidth]{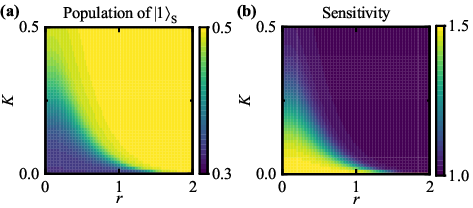}
	\caption{(a) Population at the bias point changing with nonlinear strength $K$ and parameter $r$.
		(b) Sensitivity changing with the nonlinear strength $K$ and parameter $r$.}
	\label{Fig2}
\end{figure*}

Finally, it should be emphasized that our analysis incorporates the amplification effect of the two-phonon drive on noise within the decoherence function $\chi(t)$. Consequently, the final sensitivity result aligns with those reported in Ref.~\cite{Pistolesi2021}. However, in practical experimental implementations, the mechanically amplified noise can be circumvented through a dissipative squeezing approach~\cite{Qin2018,OckeloenKorppi2017}. In this scenario, the sensitivity becomes
\begin{equation}\label{senF}
	\delta F_{\text{min}}^{\text{opt}} = \frac{\omega_R^3}{\omega_F^3} \frac{\hbar\sqrt{e\gamma_0}}{2x_0e^r},
\end{equation}
where the optimum occurs at $t = 1/\gamma_0$. Compared with traditional mechanical Fock qubit schemes~\cite{Pistolesi2021}, our mechanical squeezed qubit approach is still exponentially improved.

%
\end{bibunit}
\thispagestyle{empty}
\end{document}